\documentclass[letter,scriptaddress,twocolumn, prl,showpacs,superscriptaddress] {revtex4-1}

\usepackage[version=3]{mhchem} 
\usepackage{graphicx}
\usepackage{bm}
\pdfoutput=1

\begin{document}

\title{Suspended graphene variable capacitor}

\author{M. AbdelGhany}
\email{mohamed.abdelghany@mail.mcgill.ca}
\affiliation{Department of Electrical and Computer Engineering, McGill University, Montreal, Quebec H3A 0E9, Canada}
\author{F. Mahvash}
\affiliation{Department of Electrical and Computer Engineering, McGill University, Montreal, Quebec H3A 0E9, Canada} 
\affiliation{D\'{e}partement de Chimie et Biochimie, Universit\'{e} du Qu\'{e}bec \`{a} Montr\'{e}al, Montr\'{e}al, Qu\'{e}bec, H3C 3P8, Canada}
\author{M. Mukhopadhyay}
\affiliation{Department of Electrical and Computer Engineering, McGill University, Montreal, Quebec H3A 0E9, Canada} 
\author{A. Favron}
\affiliation{D\'{e}partement de Physique, Universit\'{e} de Montr\'{e}al, 2900 boul. \'{E}douard-Montpetit, Montr\'{e}al, Qu\'{e}bec H3C 3J7, Canada}
\author{R. Martel}
\affiliation{D\'{e}partement de Chimie, Universit\'{e} de Montr\'{e}al, 2900 boul. \'{E}douard-Montpetit, Montr\'{e}al, Qu\'{e}bec H3C 3J7, Canada}
\author{M. Siaj}
\affiliation{D\'{e}partement de Chimie et Biochimie, Universit\'{e} du Qu\'{e}bec \`{a} Montr\'{e}al, Montr\'{e}al, Qu\'{e}bec, H3C 3P8, Canada}
\author{T. Szkopek}
\email{thomas.szkopek@mcgill.ca}
\affiliation{Department of Electrical and Computer Engineering, McGill University, Montreal, Quebec H3A 0E9, Canada}

\date{\today}

\maketitle

\textbf{ The tuning of electrical circuit resonance with a variable capacitor, or varactor, finds wide application with the most important being wireless telecommunication. We demonstrate an electromechanical graphene varactor, a variable capacitor wherein the capacitance is tuned by voltage controlled deflection of a dense array of suspended graphene membranes. The low flexural rigidity of graphene monolayers is exploited to achieve low actuation voltage in an ultra-thin structure. Large arrays comprising thousands of suspensions were fabricated to give a tunable capacitance of over 10 pF/mm$^{2}$, higher than that achieved by traditional micro-electromechanical system (MEMS) technologies. A capacitance tuning of 55\% was achieved with a 10 V actuating voltage, exceeding that of conventional MEMS parallel plate capacitors. Capacitor behavior was investigated experimentally, and described by a simple theoretical model. Mechanical properties of the graphene membranes were measured independently using Atomic Force Microscopy (AFM). Increased graphene conductivity will enable the application of the compact graphene varactor to radio frequency systems.}

Mechanically tuned variable capacitance has been an effective means to tune resonant circuits since the advent of radio \cite{Korda_patent}. More compact varactors have since been developed in the form of an electrically tuned semiconductor junction capacitance \cite{Barnes_patent}. Micro-electromechanical system (MEMS) implementations of varactors \cite{Larson_1999} combine the advantages of mechanical and semiconductor varactors in a single device architecture, including high electrical quality factor, high linearity, and the capacity for monolithic integration with silicon electronics \cite{Rebeiz_book}. The canonical MEMS varactor is the parallel plate structure consisting of a conducting membrane suspended over a fixed plate, actuated by electrostatic attraction under an applied bias potential. While simple in structure, typical parallel plate varactors suffer high operating voltage\cite{Young_1997,modeling_2009} and a limited capacitive tuning range. These limitations are typically overcome by complex mechanisms \cite{Baek_2015}, which increase both the size and actuation voltage.

Fundamentally, increasing the flexibility of a suspended element by reducing it's thickness will reduce the actuation voltage of a parallel plate varator \cite{abdelghany_2012}. Monolayer graphene membranes achieve the ultimate limit with an inferred elastic stiffness of $E_{e} \simeq$ 390 N/m \cite{Bunch_2008}. In comparison a 15 nm thick Si$_{3}$N$_{4}$ membrane has an elastic stiffness of $E_{e} \simeq$ 6.3 kN/m. In addition to lower actuation voltage, graphene nano-electromecanical systems (NEMS) occupy less area than traditional MEMS counterparts\cite{chen_2013}, and can be easily integrated with integrated silicon electronics using standard transfer techniques \cite{Chen_2009,Huang_2015}. In the last decade, graphene NEMS have been widely investigated, including suspended graphene resonators\cite{Bunch_2007,song_2012,weber_2014}, switches\cite{milaniana_2009,li_2012,li_2015}, and sensors\cite{Bunch_2008}. While the theoretical limits of suspended graphene varactors has been investigated \cite{abdelghany_2012}, large arrays of low spring constant suspensions has been plagued by low fabrication yield \cite{vanderZande_2010}. In this letter we report the fabrication and characterization of suspended graphene varactors. Each fabricated device is an array of over 1000 suspensions. The total tuneable capacitance of each varactor is $C_{v} \geq 1~\mathrm{pF}$.

\begin{figure*}
\includegraphics[width=\linewidth]{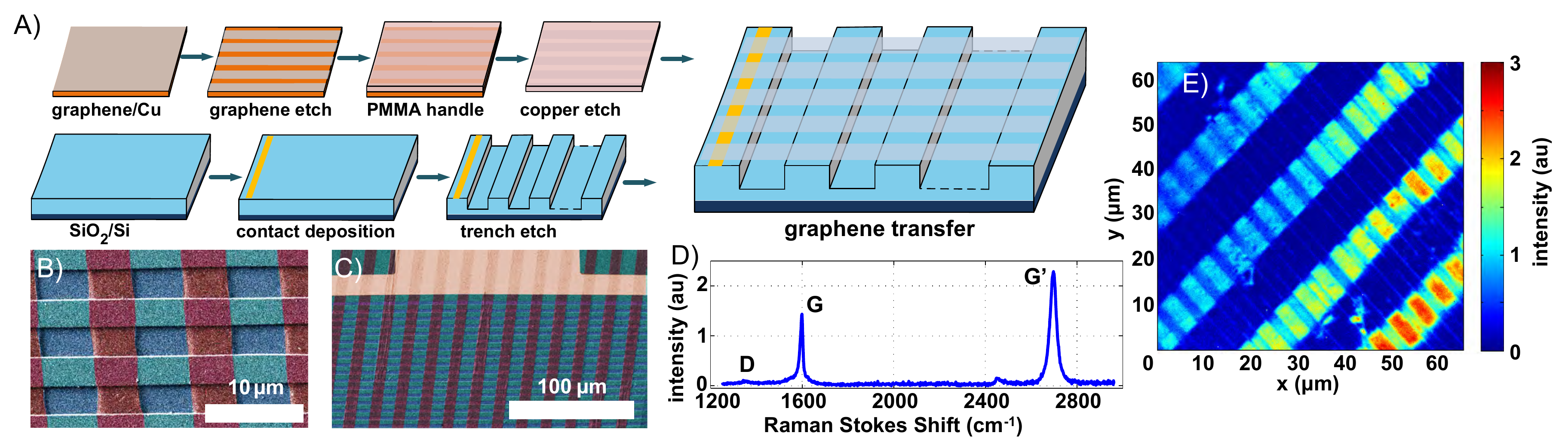}
\caption{A) The graphene varactor fabrication process includes substrate preparation, graphene patterning, and finally graphene transfer. The varactor is shown at the far right. B), C) False color scanning electron micrographs of graphene suspensions from two different varactors. D) The Raman spectrum of monolayer graphene after suspension. E) Spatial map of the Raman G' peak intensity of graphene suspensions.} 
\label{device_and_process}
\end{figure*}

The varactors were fabricated by the process illustrated in Figure \ref{device_and_process} A), consisting of three stages. The first stage is substrate preparation. Low resistivity Si wafers with 300 nm of thermal SiO$_{2}$ were metallized and trenches were etched. The second stage is graphene growth and pre-patterning. Growth on Cu foils was performed by chemical vapor deposition, followed by photolithography and dry etching of graphene strips, and lastly a polymethylmethacrylate (PMMA) handle was used during a sacrificial Cu etch. In the final stage, graphene strips were transferred atop the trenches in a wet process with the suspensions released by critical point drying. Further processing details are provided in the supplementary information. A schematic of the fabricated device is illustrated in Figure \ref{device_and_process}A). Trenches of depth $h=155~\mathrm{nm}$ depth and length $L=2.5~\mathrm{\mu m}$ were fabricated, in arrays of at least 20 by 50 suspensions with an areal density of capacitance of $12~\mathrm{pF/mm}^{2}$. The active area of graphene suspensions constitutes $20\%$ of the total device area.

Figures \ref{device_and_process}B) and \ref{device_and_process}C) are false colour scanning electron micrographs of two different devices. Importantly, a layer of SiO$_{2}$ at the bottom of the trench ensures that a single suspension collapse does not result in a short circuit. A critical parameter determining device yield is the aspect ratio of the trench length $L$ to the trench depth $h$. Previously reported yields of suspended graphene arrays were $>80\%$ for suspended strips over trenches with aspect ratios $L/h < 10$ \cite{vanderZande_2010}, and $>90\%$ for circular resonators over holes with aspect ratios less than 6.7\cite{barton_2011}. We have achieved a yield of $\geq 95 \%$ for devices with aspect ratio $L/h > 16$.

\begin{figure}
\includegraphics[width=\linewidth]{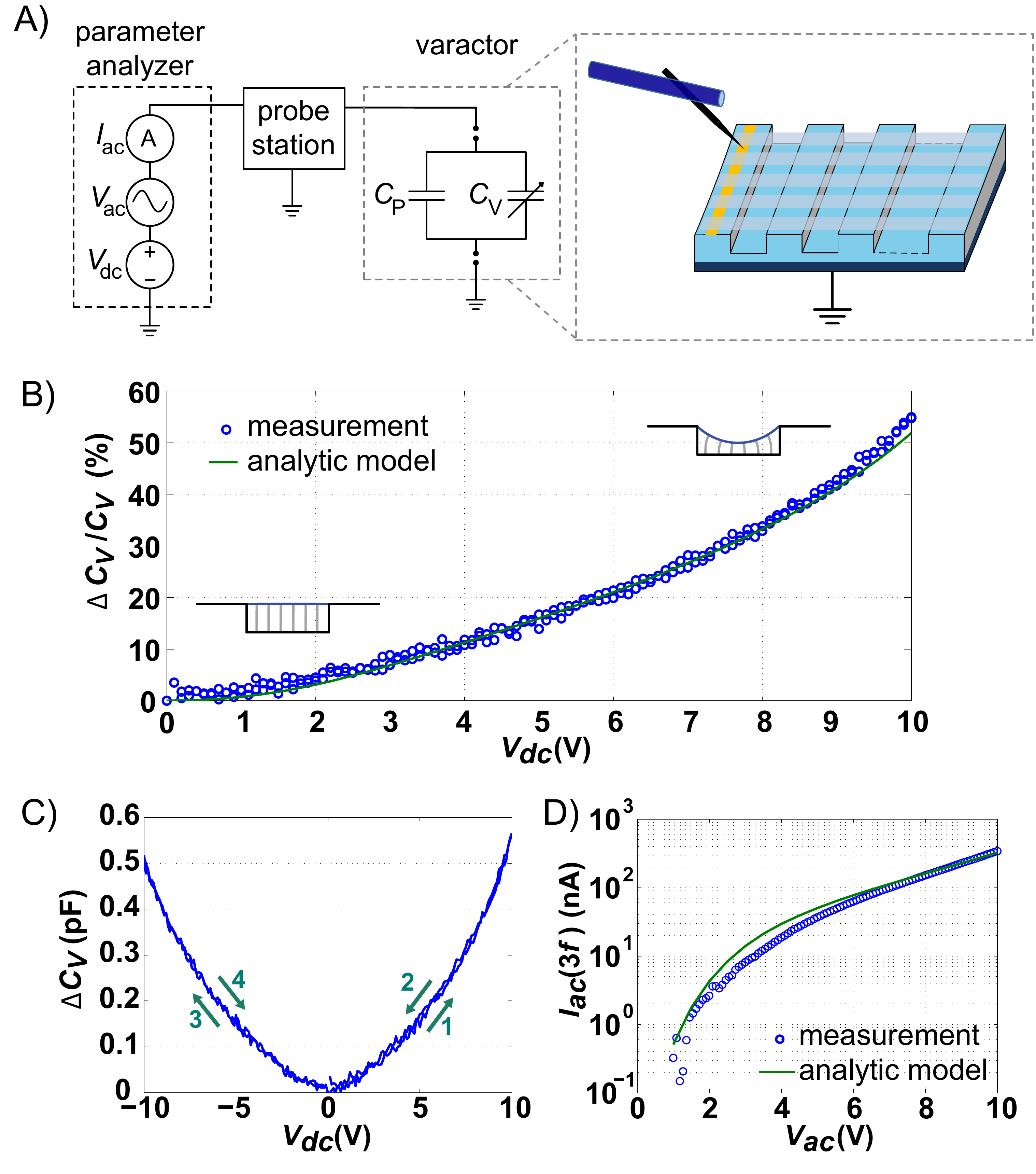}
\caption{A) The schematic of the capacitance measurement setup in a vacuum probe station, including both variable and parasitic capacitance of the varactor, $C_V$ and $C_P$, respectively. B) The relative change in capacitance $[C_V(V_{dc})-C_V(0)]/C_V(0)$ versus bias $V_{dc}$ as measured with $V_{ac}=30~\mathrm{mV}$ at $f=100~\mathrm{kHz}$ and best-fit to a simple analytical model. The best-fit model parameters are $S_0 = 40~\mathrm{mN/m}$ and $E = 180~\mathrm{GPa}$. C) The change in varactor capacitance $[C_V(V_{dc})-C_V(0)]$ versus $V_{dc}$, illustrating ambivalent operation. The measurement shows forward and backward sweeps with no hysteresis. D) The third harmonic current $I_{ac}(3f)$ versus $V_{ac}$ as measured and with a simple analytic model with no fit parameters. The modulus $E$ and pre-tension $S_0$ were determined from $C_V$ versus $V_{dc}$. } 
\label{setup_and_result}
\end{figure}

Raman spectroscopy was used to confirm the quality of the suspended graphene. Figure \ref{setup_and_result}D) shows a representative Raman spectrum with a high intensity G' peak ($2697~\mathrm{cm}^{-1}$) and a sharp G peak ($1598~\mathrm{cm}^{-1}$) indicative of predominantly monolayer graphene \cite{raman_book}. The peak intensity ratio of the defect related D band ($1353~\mathrm{cm}^{-1}$)) to the G peak is $I_{D}/I_{G}$ is 0.077, indicative of a low defect density \cite{raman_book}. Furthermore, hyperspectral Raman imaging was used to independently confirm the continuity of graphene. Figure \ref{device_and_process}E) shows a spatial Raman map of the G' peak intensity over a $65~\mathrm{\mu m} \times 65~\mathrm{\mu m}$ area. The peak intensity of supported graphene is greater than suspended graphene as the oxide thickness is optimal for Fabry-P\'{e}rot enhancement of optical intensity. 

The varactor capacitance was measured in a vacuum probe station, as schematically shown in Figure \ref{setup_and_result}A), including both the fixed parasitic capacitance $C_P~15~\mathrm{pF}$ and tuneable capacitance $C_V$. A semiconductor parameter analyzer was used to measure the capacitance with an ac excitation of $V_{ac}=30~\mathrm{mV}$ at a frequency $f=100~\mathrm{kHz}$ while a dc bias voltage $V_{dc}$ was swept to tune capacitance by electrostatic actuation of the suspended graphene membranes.

Figure \ref{setup_and_result}B) shows the typical behavior of tuned capacitance $\Delta C_V / C_V$ versus bias $V_{dc}$ for a typical graphene varactor among the five devices tested. At a bias voltage $V_{dc}=10~V$, the capacitance change is $55 \%$, exceeding the $50 \%$ pull-in limit of a Hookean parallel plate varactor \cite{modeling_2009}. The measurement results agree well with a virtual displacement model imposing a balance of stretching, pre-tension and electrostatic forces on a graphene membrane, \begin{equation}
\dfrac{Et\pi^{5}\delta^{3}}{8L^{4}(1-\nu^2)} + \frac{8C_{1}S_{0}\delta}{L^{2}} =\frac{ \epsilon V_{dc}^{2} } { \left( h'   - \sqrt{\frac{2}{\pi}} \delta  \right)^{2} },
\label{eq:uniform}
\end{equation}
where the membrane shape is approximated with a half-cosine, $E$ is the graphene Young modulus, $S_0$ is the pre-tension, $t=3.35~\AA$ is the graphene thickness, $\nu=0.141$ is the graphene Poisson raio, $C_{1}=2$ is a constant dependent on membrane aspect ratio \cite{vlassak_1992}, $h'=192~\mathrm{nm}$ is the electrical length between Si substrate and graphene membrane, and $\delta$ is the graphene membrane deflection induced by applied bias $V_{dc}$. The capacitance $C_V$ is simply expressed in terms of the deflection \cite{abdelghany_2012},
\begin{equation}
C_V = \frac{4WL\epsilon_0}{\pi h' \sqrt{1-\delta^2/h'^2)} }\arctan \left( \sqrt{ \frac{ h' + \delta } {h' - \delta} } \right)
\end{equation}
Fitting the experimentally measured capacitance versus bias voltage, we extract a Young's modulus $E =180~\mathrm{GPa}$ and a pretension $S_{0}=40~\mathrm{mN/m}$. Both $E$ and $S_0$ are lower than that reported in experiments with individual exfoliated graphene membranes \cite{Bunch_2007}.

The ambivalent response of the varactor to the applied bias $V_{dc}$ was measured, as shown in Figure \ref{setup_and_result}C). Low hysteresis was observed for a 10~V swing of $V_{dc}$ of either polarity. The non-linearity of the ambivalent response leads to odd harmonic generation in the resulting current spectrum, since the total ac current $I_{ac}=CdV/dt + VdC/dt$, where $dC/dt=dC/d\delta \cdot d\delta/dt$. We measured the third harmonic current $I_{ac}(3f)$ with a lock-in amplifier tuned for ac excitation $V_{ac}=1~\mathrm{V}$ to $10~V$ at a frequency of $f=20~\mathrm{kHz}$. The measured $I_{ac}(3f)$ versus $V_{ac}$ is plotted in Figure \ref{setup_and_result}D). A simple model for $I_{ac}$ employing a quasi-static approximation for $d\delta/dt$ ignoring inertial effects leads to excellent agreement with measured results without any free fitting parameters. Notably, at $V_{ac} =1~\mathrm{V}$ the ratio $I_{ac}(f)/I_{ac}(3f)=220$, corresponding to less than$-46~\mathrm{dB}$ third harmonic distortion in the varactor response.

\begin{figure}
\includegraphics[width=\linewidth]{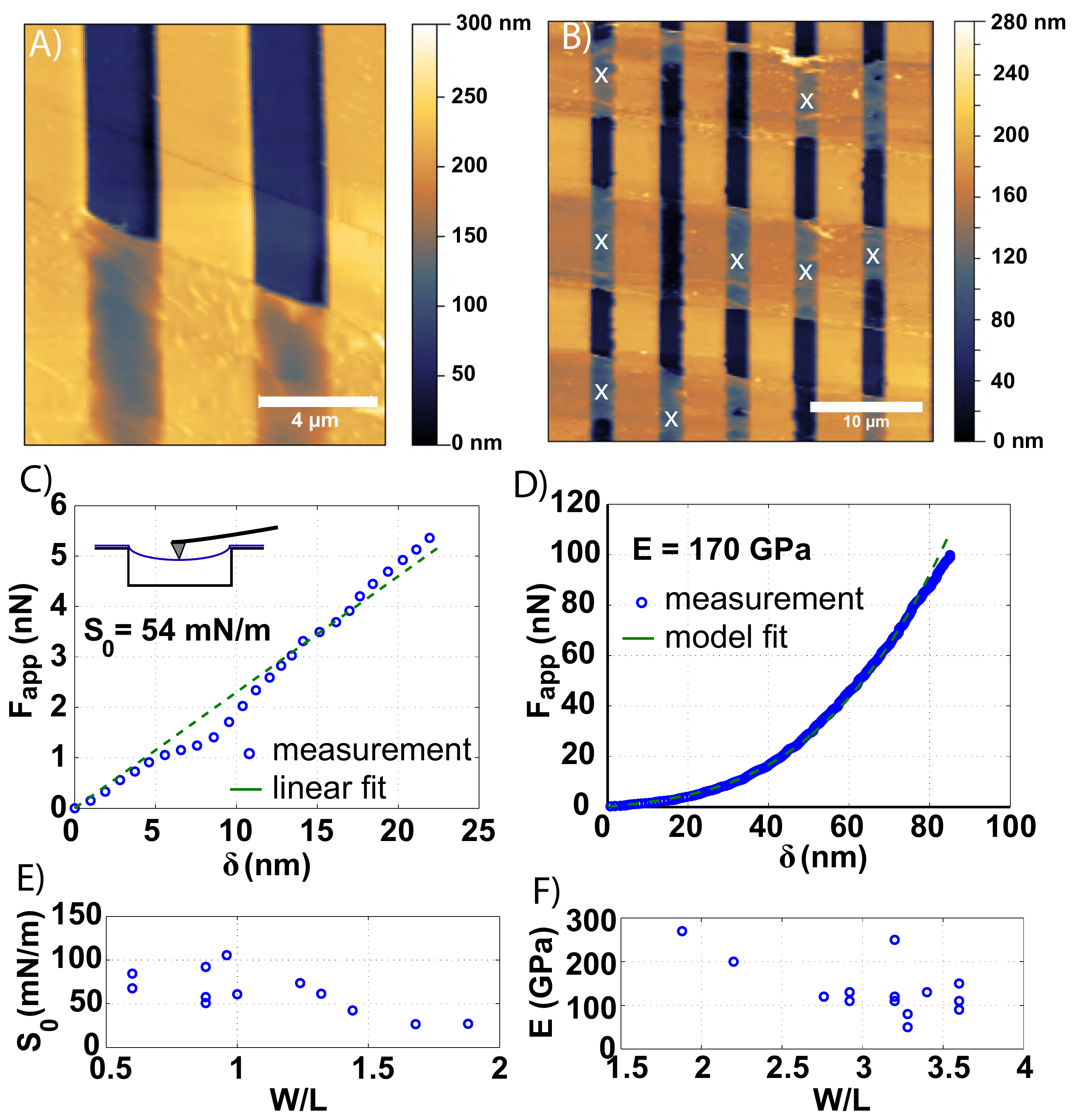}
\caption{A) Contact mode AFM image of a portion of a graphene varactor. B) Contact mode AFM image of a graphene varactor after high voltage stress testing, illustrating membranes that are in tact, membranes that have partially collapsed, and membranes that have completely collapsed. Crosses indicate locations where force-deflection measurements were taken. C) Applied force $F_{app}$ versus graphene deflection $\delta$ as measured by AFM, and with a linear fit appropriate to Hookean response. The inferred pre-tension is $S_{0} = 54~\mathrm{mN/m}$. D) Applied force $F_{app}$ versus graphene deflection $\delta$ at large deflection, as measured by AFM and with a model fit. The inferred Young's modulus from the model is $E=170~\mathrm{GPa}$. E) Summary of pre-tension $S_0$ inferred from linear deflection versus membrane $W/L$ aspect ratio for 12 membranes. F) Summary of inferred Young's modulus $E$ inferred from non-linear deflection versus membrane $W/L$ aspect ratio for 14 membranes. } 
\label{afm}
\end{figure}

Atomic force microscopy (AFM) was used to independently verify the mechanical properties of individual suspended graphene membranes. Contact mode AFM images were first taken of the varactor, as shown in Figure \ref{afm}A) and B). Force-deflection measurements\cite{frank_2007} were then conducted on 33 individual membranes, several of which are indicated in Figure \ref{afm}B). A variety of membranes were selected, including several that underwent partial collapse following high voltage testing of varactor response. The deflection of a silicon cantilever with a calibrated spring constant ($k_{cant}=0.916~\mathrm{N/m}$) was measured as a function of the piezoelectric driven extension of the AFM, from which the force-displacement curve was inferred. Figure \ref{afm}C) depicts a typical measurement of applied force $F$ versus graphene membrane displacement $\delta$. For deflections $\delta < 25~\mathrm{nm}$, a linear fit determines the effective spring constant $k_{graphene}$ per Hooke's law. Pre-tension $S_0$ dominates in the linear regime, and was estimated according to the beam approximation under a point load, $k_{graphene} \simeq (\pi^2/2) S_{0}W/L$. The pre-tension $S_0$ of 12 membranes are illustrated in Figure \ref{afm}E) versus $W/L$. Non-linearity in applied force versus deflection was observed at larger deflections due to graphene stretching, with an example illustrated in Figure \ref{afm}D). The force versus deflection relation $F\propto E \delta^3$ can be modelled for different geometries with a virtual displacement method (details provided in supplementary information), allowing a numerical fit to measurements and extraction of the Young's modulus $E$. The Young's modulus $E$ of 14 membranes are illustrated in Figure \ref{afm}F) versus $W/L$, yielding a mean $E=140\pm60~\mathrm{GPa}$ in good agreement with the value $E=170\pm5~\mathrm{GPa}$ determined from capacitance measurements of the same varactor.

 \begin{figure}
 \includegraphics[width=\linewidth]{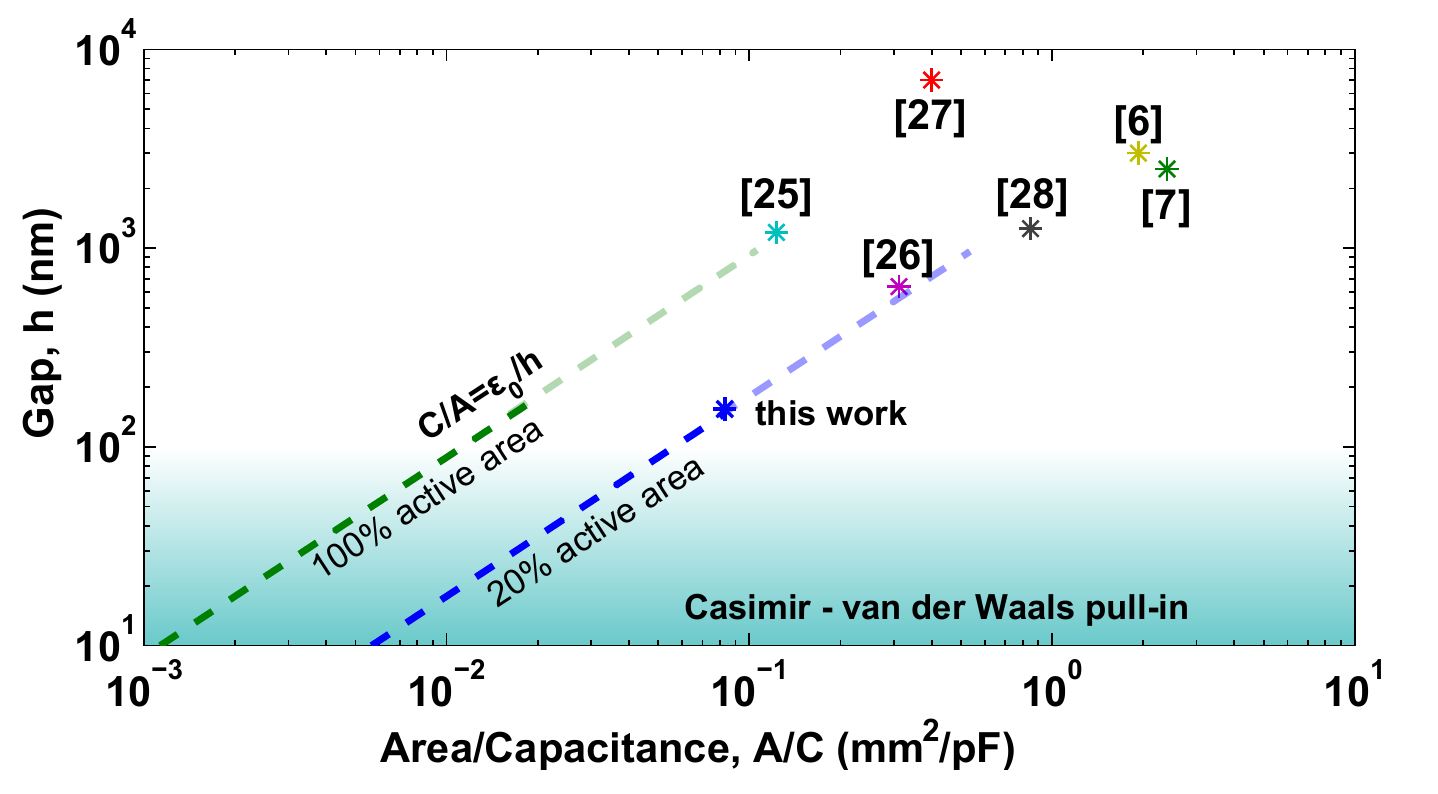}
 \caption{A comparison between this work and different state of the art MEMS varactors, with varactor gap $h$ and area per unit capacitance $A/C$ compared. The devices shown are state of the art silicon based MEMS varactors: with fractal structure\cite{elshurafa_2013}, curved plate varactor\cite{kassem_2008}, vertical parallel plate\cite{achenbach_2014}, parallel plate with levers\cite{levers_2011}, simple parallel plate design\cite{modeling_2009}, a comb finger varactor\cite{Baek_2015}. Further indicated in the graph are the ideal limits achievable in accordance with $C/A = \epsilon_0/h$. The varactor height is limited by spontaneous pull-in from Casimir-van der Waals forces.}
 \label{comparison}
 \end{figure}
 
We finally consider a comparison of varactor height and area required to achieve $C_V = 1~\mathrm{pF}$, including the suspended graphene varactor and various state of the art MEMS varactors. The suspended graphene varactor offers the highest capacitance density in the smallest vertical space. Improving the active area occupation of total device area beyond $20\%$ will further increase the capacitance density of the graphene varactor. At $100\%$ active area occupation, the areal capacitance density reaches the limit imposed by the permittivity of free space, $C/A = \epsilon_0/h$. The ultimate limit to the achievable areal capacitance density of suspended graphene will be determined by the minimum trench depth $h$ that can be sustained without spontaneous pull-in by Casimir-van der Waals forces. The criterion for spontaneous pull-in by Casimir-van der Waals forces\cite{serry_1998} in the ideal conductor limit is $R L^4 /E t h^7 < 0.245$ with $R=\hbar c \pi^2/240$. For a trench aspect ratio $L/h = 10$, a minimum trench height of $h=10~\mathrm{nm}$ and maximum capacitance density of $C/A = 890~\mathrm{pF/mm}^2$ can be theoretically achieved.
 
In conclusion, we have demonstrated large area suspended graphene varactors, reaching a $55\%$ tuning range with a 10~V actuation voltage, a high device yield $\geq 95\%$ and an areal capacitance density of $12~\mathrm{pF/mm}^{2}$. Further reduction in pull-in voltage may be achieved by increasing the trench aspect ratio $L/h$, but avoiding spontaneous pull-in by Casmir-van der Waal forces will require increased height $h$, and thus reduced areal capacitance. The application of graphene varactors to radio frequency circuits requires the challenge of monolithic integration to be addressed through more advanced fabrication processes.

\onecolumngrid
\appendix

\newpage
\pagebreak

\renewcommand{\theequation}{S\arabic{equation}}
\setcounter{equation}{0}
\renewcommand{\thefigure}{S\arabic{figure}}
\setcounter{figure}{0}
\renewcommand{\bibnumfmt}[1]{[S#1]}
\renewcommand{\citenumfont}[1]{S#1}
\linespread{2}

\section{Supplementary information for Suspended graphene variable capacitor}
\begin{center}
M. AbdelGhany$^\ast$, F. Mahvash, M. Mukhopadhyay, A. Favron, R. Martel, M. Siaj, and T. Szkopek$^\dagger$
\end{center}

\subsection{Fabrication Process}
\subsubsection{Substrate preparation}
The devices were fabricated on a low resistivity silicon wafer $(\rho \simeq 0.005 \mathrm{\Omega.cm})$ with $300$ nm of thermal oxide on both sides. The oxide was completely removed from one side to allow access to the silicon. It was removed by wet itching in a 10:1 diluted hydrofluoric acid $(HF)$ for 20 minutes, while covering the other side of the wafer with photo-resist and protective tape to preserve the oxide on this side. Afterwards metal contacts were deposited on the front side using lift-off process. Photolithography was used for lift-off, we spun 700 nm of LOR 5B lift-off resist under 1.4 $\mu m$ of S1813 positive resist to create an undercut for lift-off. The metal was deposited using electron beam evaporation. Metal contacts consisted of 100 nm gold over 10 nm titanium, the titanium improved the adhesion of the contacts. The lift-off was done by putting the wafer in  Remover 1165 at $70^{\circ}C$ with sonication for 20 minutes. Afterwards the wafer was transferred to a fresh beaker of  Remover 1165 and left for another 10 minutes at $70^{\circ}C$, then the wafer was rinsed in Isopropyl alcohol (IPA) and deionized (DI) water for five minutes each. Oxygen plasma was used to get rid of all resist residue.

The trenches were etched using photolithography and reactive ion etching (RIE). The etch rate was calibrated for narrow trenches as it decreases with trench width. Atomic force microscopy (AFM) was used to accurately measure the trench depth and calibrate the etch rate. After the desirable depth was achieved, the resist was removed using the same method described above.

\subsubsection{Graphene growth and pre-patterning}
\begin{figure}
	\includegraphics[width=\linewidth]{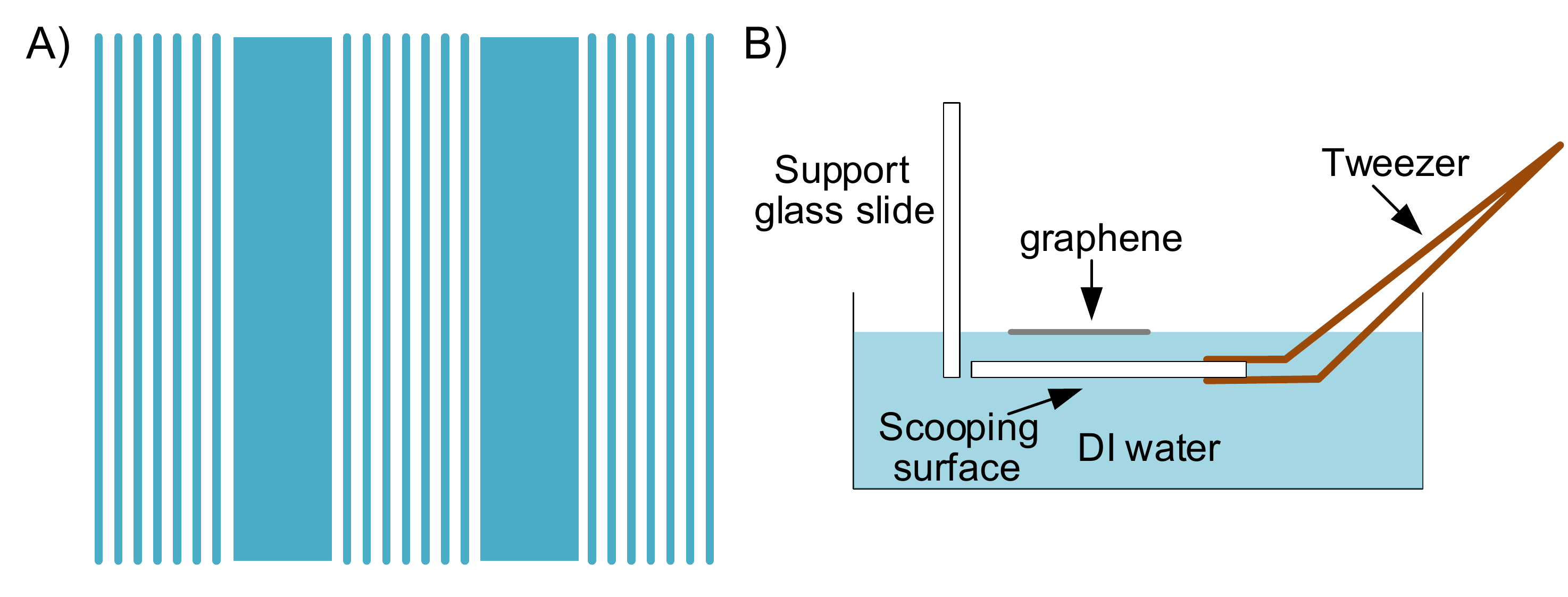}
	\caption{A) Part of the mask used for patterning graphene. Continuous pieces of graphene are used as alignment marks. B) A diagram showing how the $PMMA$ supported graphene is scooped out of a liquid} 
	\label{fig_S1}
\end{figure}

We used large area graphene grown by chemical vapor deposition (CVD). We grew single layer graphene on both sides of a 25 $\mu m$ sheet of copper foil. Before transfer the graphene on the desired side was pre-patterned using photolithography and RIE to create graphene strips to facilitate suspension as well as alignment marks to help transfer the graphene strips orthogonal to the trenches. Figure \ref{fig_S1}A) shows the graphene pattern used. The continuous pieces acted as alignment marks, as there is a clear difference in transparency between them and the regions with narrow strips.
The graphene was etched using oxygen plasma RIE with 100 W RF power, 200 mT pressure, and a gas flow of 40 scc. After the patterning, the copper pieces were left in acetone for five minutes to remove the photoresist, then rinsed in IPA and DI water for five minutes each. 

\subsubsection{Graphene transfer}
To prepare for transfer, The patterned graphene was covered with 300 nm of PMMA 950 A4 polymer handle for mechanical support. The polymer was then baked at $90^{\circ} C$ for three minutes. The copper was etched in a 0.1 molar solution of ammonium per-sulphate $(NH_4)_2S_2O_2$, after $45 \sim 60$ minutes of etching the samples were removed and the back side of the copper foil was sprayed with DI water to remove the graphene on that side. After the copper etching was done (in $18 \sim 24$ hours) the graphene with the polymer handle was left floating over the $(NH_4)_2S_2O_2$. It was then scooped from the etchant and transferred to a beaker of DI water and left for five minutes. After that the graphene was transferred to a fresh beaker of DI water to get rid of all $(NH_4)_2S_2O_2$ residue. Finally the graphene is scooped on the prepared substrate and left to dry for $\sim 24$ hours. Figure \ref{fig_S1}B) shows how the graphene was scooped out of liquids.

After the sample dried, the PMMA was removed by putting the sample in acetone for four hours, then transferring it to a fresh beaker of acetone for 30 minutes. The sample was then transferred to a beaker of IPA and left for five minutes. It was transferred to a fresh beaker of IPA two or three more times to get rid of all acetone residue. While transferring the graphene the sample had to be kept horizontal so that it would be covered in liquid at all time, this was crucial to prevent the collapse of suspended strips. The sample was then transferred in the same manner to the chamber of an 'Automegasamdri®-915B, Series B' critical point dryer. For our $2 \mathrm{cm} \times 2 \mathrm{cm}$ samples, only one fourth of the chamber was use. It was filled  with enough IPA to cover the sample. The purge time was adjusted to 20 minutes. After drying, the graphene strips were cut around each individual device to separate the devices from each other. A profilometer stylus was used for cutting the graphene.

\subsection{Raman verification}
\begin{figure}
	\includegraphics[width=\linewidth]{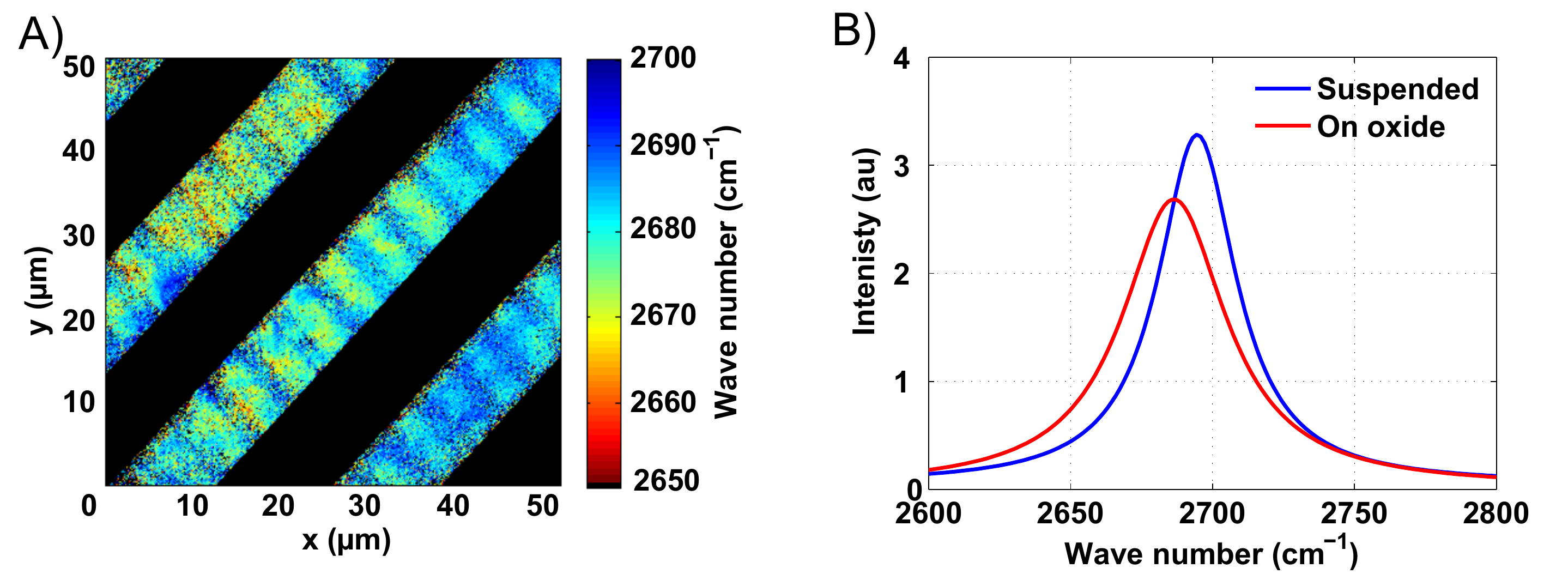}
	\caption{A) A map of the position of the G' peak, the black color donates no peak. B) Lorentzian fit for the G' peaks of two spot Raman measurements taken on the same graphene strip.} 
	\label{fig_S2}
\end{figure}

A Raman image was taken for the graphene varactor to verify the continuity of the graphene over the whole device. The image was taken using $RIMA$ hyper spectral imaging system (Photon Etc.) with pump wavelength of 532 nm. A map of the wavenumber shift of the G' peak at each spatial point is shown in figure \ref{fig_S2} A). The map illustrates that graphene strips are continuous over the trenches, with a different wavenumber shift of the G' peak between the suspended graphene and that supported on oxide. There is a blue shift in the suspended graphene. This shift was verified using spot Raman measurements, figure \ref{fig_S2} B) depicts the Lorentzian fit of G' peaks of two spots on the same strip, one suspended and the other supported. The spot measurement shows a shift of 8 $\mathrm{cm}^{-1}$ between the peaks which agrees with the hyperspectral map.
\subsection{Calibration of test equipment}
\begin{figure}
	\includegraphics[width=\linewidth]{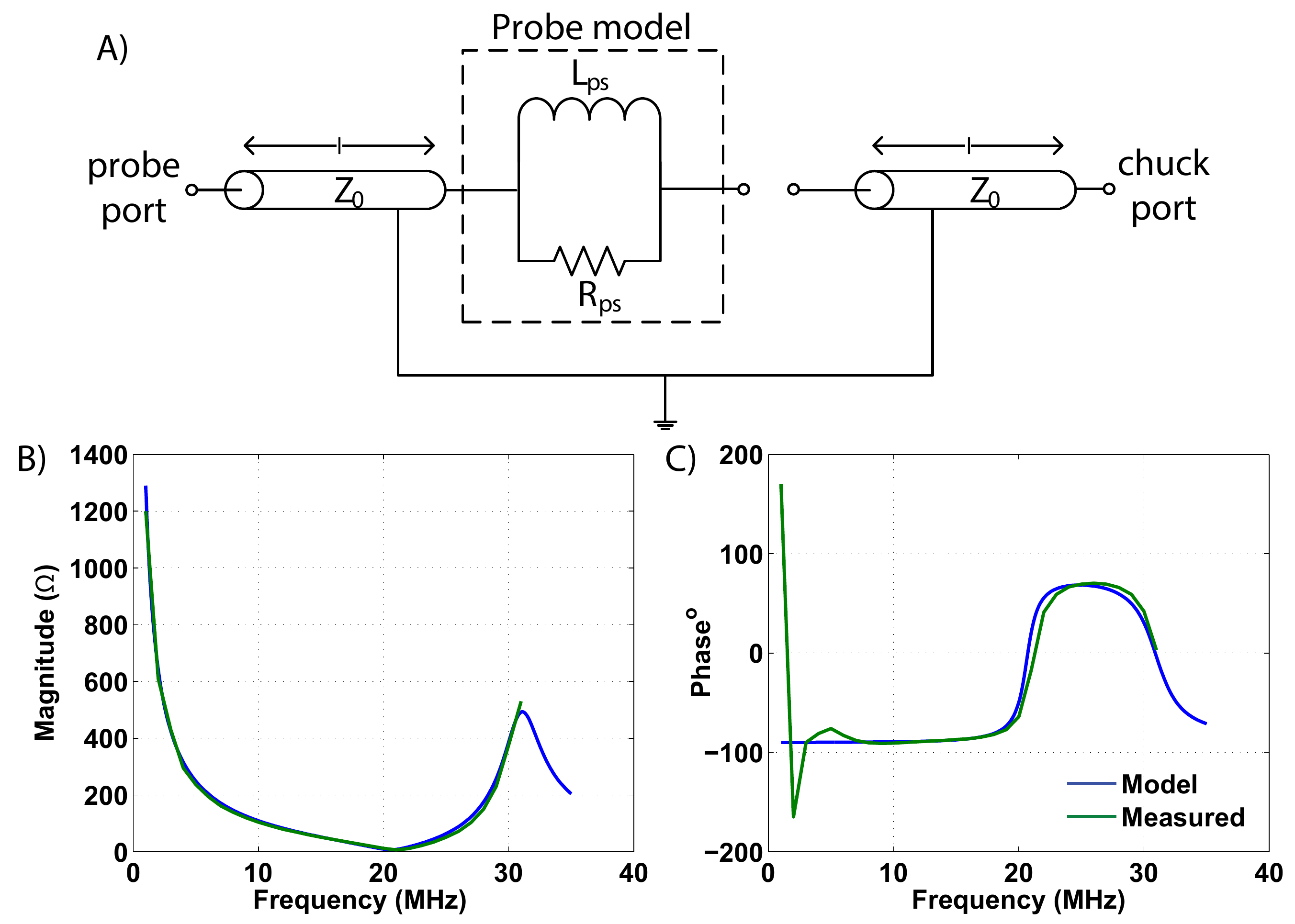}
	\caption{A) The circuit schematic of extracted probe station model. B) and C) The measured and modeled impedance of the probe station with the probe shorted to	the chuck} 
	\label{fig_S3}
\end{figure}

The electrical measurements were done in a Janis Research ST-500 probe station. The graphene was contacted by landing a probe on the metal pad, while the bulk of the silicon was fixed to the probe station chuck and acted as the other electrode. To obtain accurate results the probe station frequency response was extracted. A network analyser was used to find the S-parameters of the probe station in open-circuit and short circuit. A circuit model was extracted from these measurements. Figure \ref{fig_S3} A) depicts the extracted circuit model, while figure \ref{fig_S3} parts B) and C) compare the input impedance of the extracted model to that measured using the network analyser in the short-circuit case.

The non-linear current response of the varactor was measured using Zurich Instruments HF2 lock-in amplifier with the HF2TA current amplifier. To account for the current amplifier effects, its frequency response was measured using a known load. 

\subsection{Theoretical calculation of varactor non-linearity}
\begin{figure}
	\includegraphics[width=\linewidth]{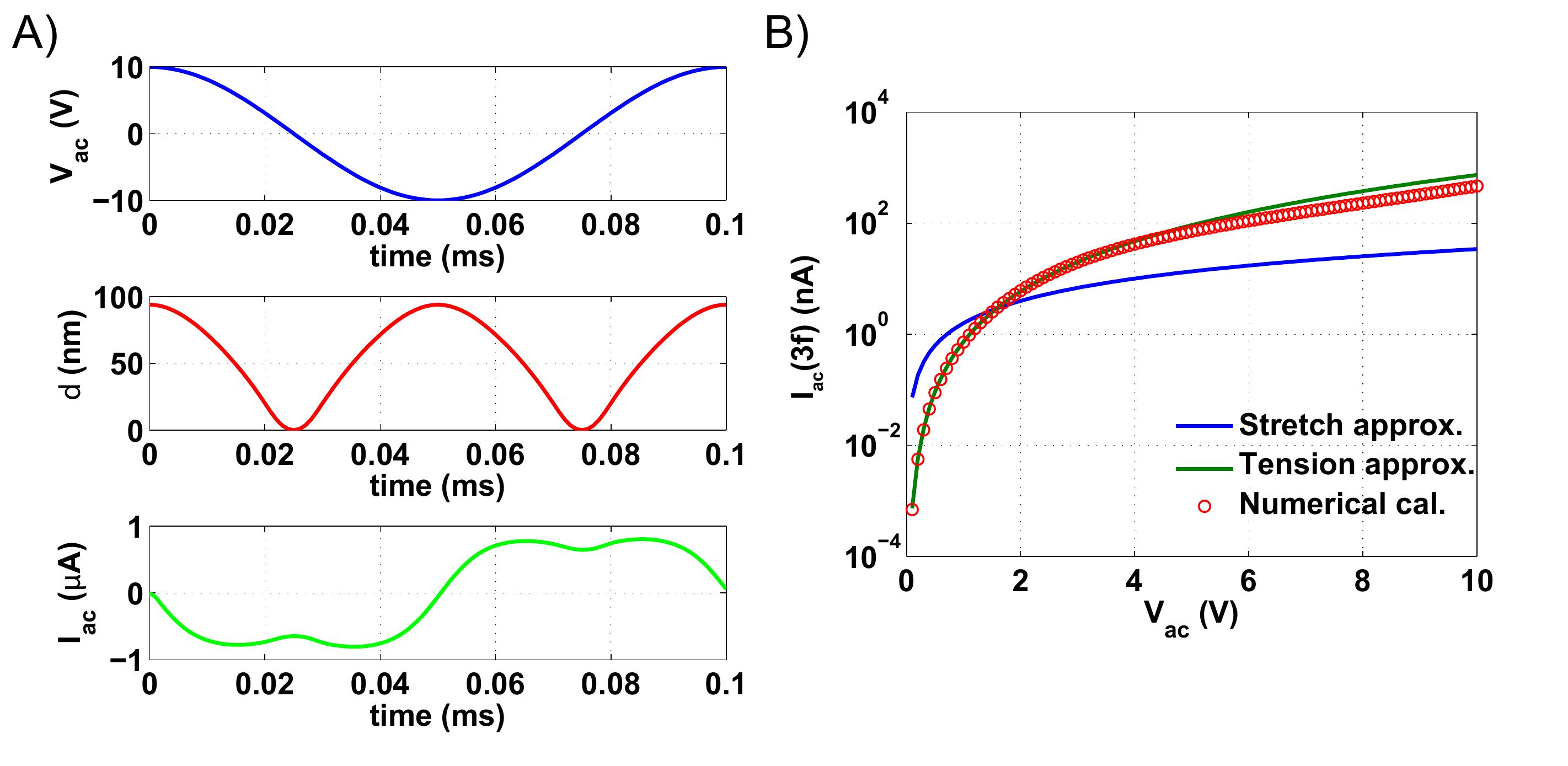}
	\caption{A) First panel: Applied AC voltage, second panel: simulated deflection of the membrane centre, third panel: resulting current. The membrane geometry extracted from SEM images of the varactor was used for this simulation, with trench length of $2.45 \mu m$ and depth of $155 nm$. The simulated AC signal has an amplitude of 10 V and a frequency of 100 kHz. B) Comparison between the third harmonic component of the varactor current estimated using the three different approximations} 
	\label{fig_S5}
\end{figure}

 Due to the complexity of the system no closed form relation could be reached for the non-linear components of the current, however numerical solutions were obtained. To facilitate the construction of a numerical model predicting the non-linearity of the varactor, it was assumed that the deflection of the suspended membranes it comprises is adiabatically invariant therefore the system is at equilibrium at all points in time and effects of damping can be ignored. This way the system was solved at each time point independently. This assumption is valid for frequencies much lower than the membrane resonant frequency, thus we limited the non-linearity study to frequencies below 1 MHz while the expected resonant frequency of the membrane is 73 MHz. Figure \ref{fig_S5}A) shows the deflection at the centre of the suspended graphene membrane and the resulting current versus time due to applied AC voltage.
For deflections much smaller than the trench depth $(d<<h)$, a closed form relation can reached. At these deflections the membrane behavior is dominated by either the stretching restoring mechanism or the residual tension restoring mechanism. In the first case the additional current due to membrane deflection is given by equation \ref{eq:Monorina}, while in the case of pretension dominated behavior, the additional current is given by equation \ref{eq:not_Monorina}. Figure \ref{fig_S5}B) shows how the two cases compares with our numerical calculations. The pretension approximation agrees with the numerical calculations for low voltage as our devices behavior is dominated by pretension at small deflection values.
\begin{equation}
I_d = C_0 K_1 \frac{L^{4/3}}{h^{5/3}}\times V^{2/3} \frac{dV}{dt}
\label{eq:Monorina}
\end{equation}

\begin{equation}
I_d = C_0 K_2 \frac{L^2}{h^3}\times V^2 \frac{dV}{dt}
\label{eq:not_Monorina}
\end{equation}
where $C_0$ is the initial capacitance, $K_1$ is a constant depending on material properties (Young's modulus and Poisson's ratio), $K_2$ is a constant depending on the membrane shape and residual tension, and  $L$ and $d$ are the capacitor length and height respectively.

\subsection{Force displacement measurement}
\begin{figure}
	\includegraphics[width=\linewidth]{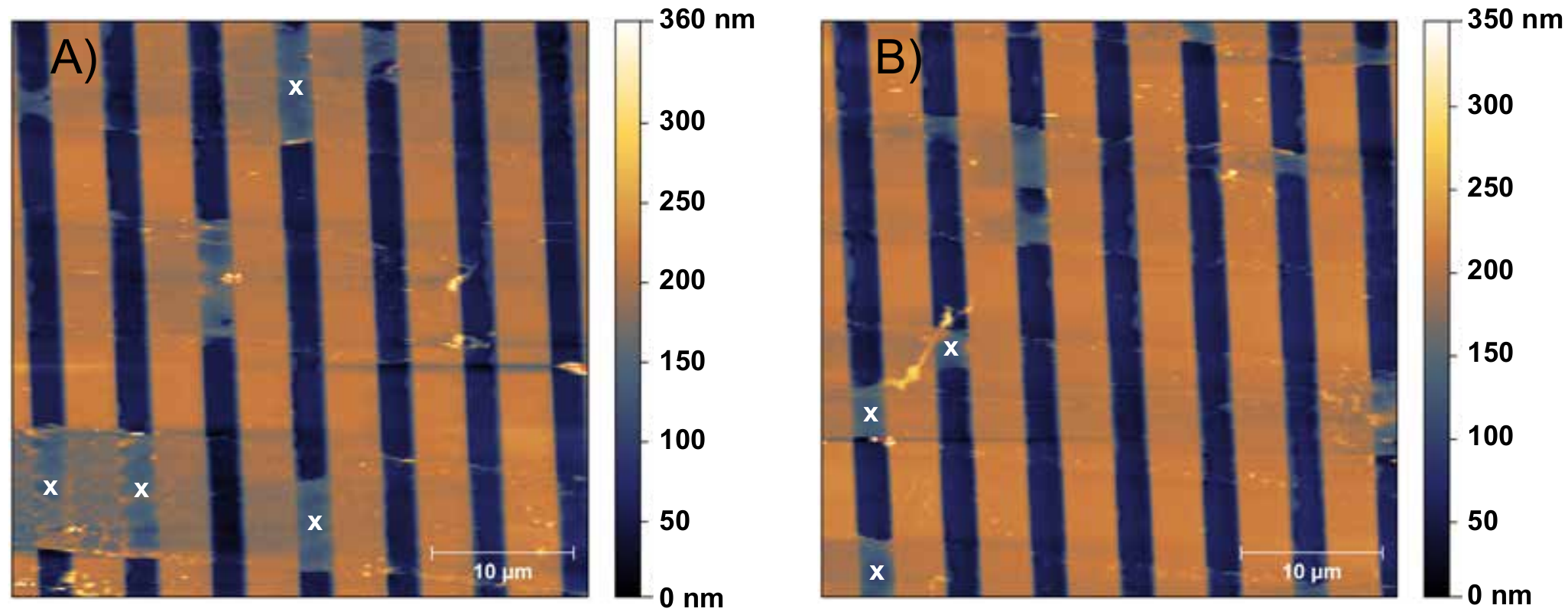}
	\caption{AFM images of two separate regions probed, the white marks show the points where the indentation was made.} 
	\label{fig_S6}
\end{figure}

The methods used in this experiment are similar to those used in references \cite{frank_2007,Lee_2008,Whittaker_2006}. The force displacement data was acquired using an Asylum MFP3D AFM. Bruker MLCT-F tips were used for the measurements, the cantilevers carrying the tips had spring constants between 0.9 N/m and 0.95 N/m. The average width of the suspended strips was 9 $\mu m$ with 90\% of the strips between 8 $\mu m$ and 9.5 $\mu m$. Thirty three suspension with widths between 1.5 $\mu m$ and 9 $\mu m$ were probed. The device was imaged before starting the experiment to accurately choose the indentation position. The geometric centre of each suspension was chosen for probing as indicated by the white marks in figure \ref{fig_S6}. Suspensions with no defects were chosen.

\begin{figure}
	\includegraphics[width=\linewidth]{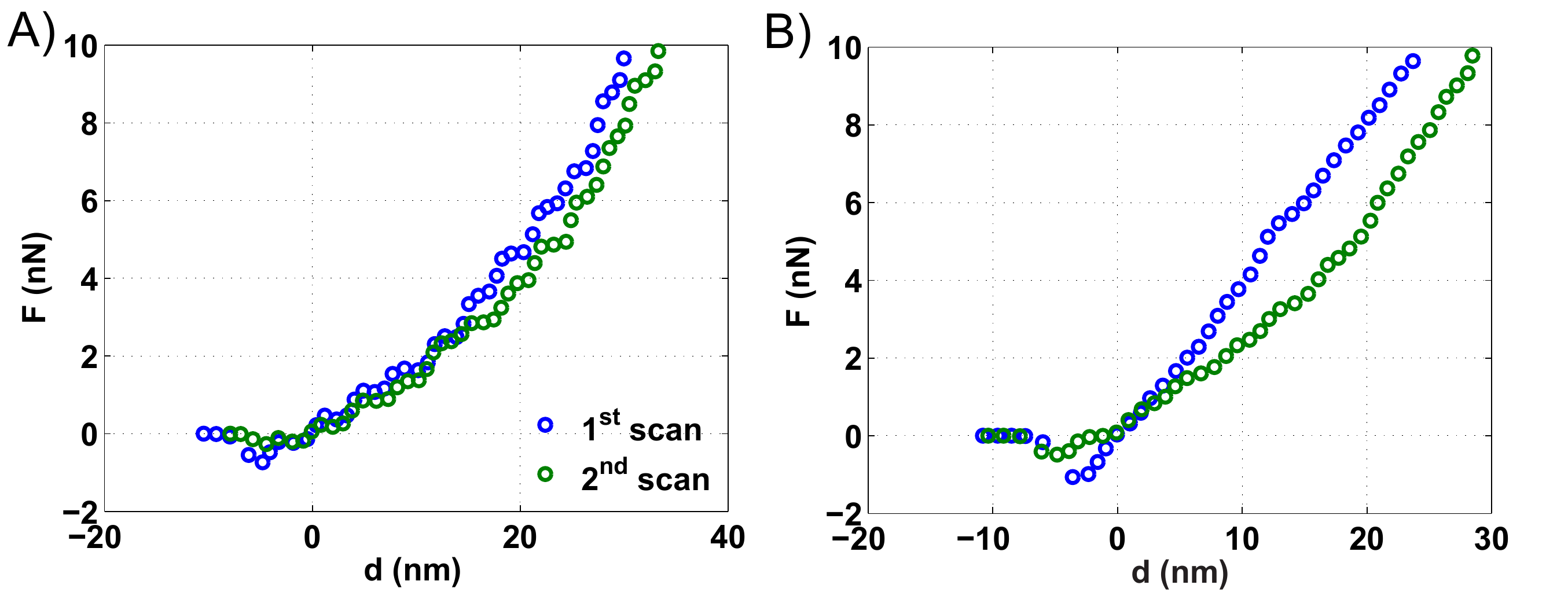}
	\caption{AFM force versus displacement curves of two suspensions. The suspension in part A) shows no signs of damage, while the suspension in part B) seems to be damaged after the first scan, and thus it was excluded from the experiment.} 
	\label{fig_S7}
\end{figure}

The microscope was kept scanning for 20 minutes before the indentation to minimize the x-y drift. Each suspension was then tested with maximum force of 10 nN. The test was repeated twice to account for slippage and breakage of the membrane. Figure \ref{fig_S7}A) shows the raw data from a suspension where the first indentation did not cause damage thus the two sets of data agree, while figure \ref{fig_S7}B) shows the data from a suspension where the first indentation caused some damage therefore the second set of data is different. Suspensions that showed signs of damage were excluded from the experiment. Half of the suspensions were then tested with maximum force of 150 nN to probe the non-linear membrane (stretching) behavior, from which Young's modulus can be extracted. The experiment recorded the piezo displacement $(\Delta Z)$, the tip deflection $(\delta_{tip})$, and the applied force $(F)$. The graphene deflection $(d)$ was extracted by subtracting the tip deflection from the piezo displacement\cite{frank_2007,Lee_2008} $d = \Delta Z - \delta_{tip}$. It was necessary to determine the point at which tip deflection is zero to obtain accurate results.

\begin{figure}
	\includegraphics[width=\linewidth]{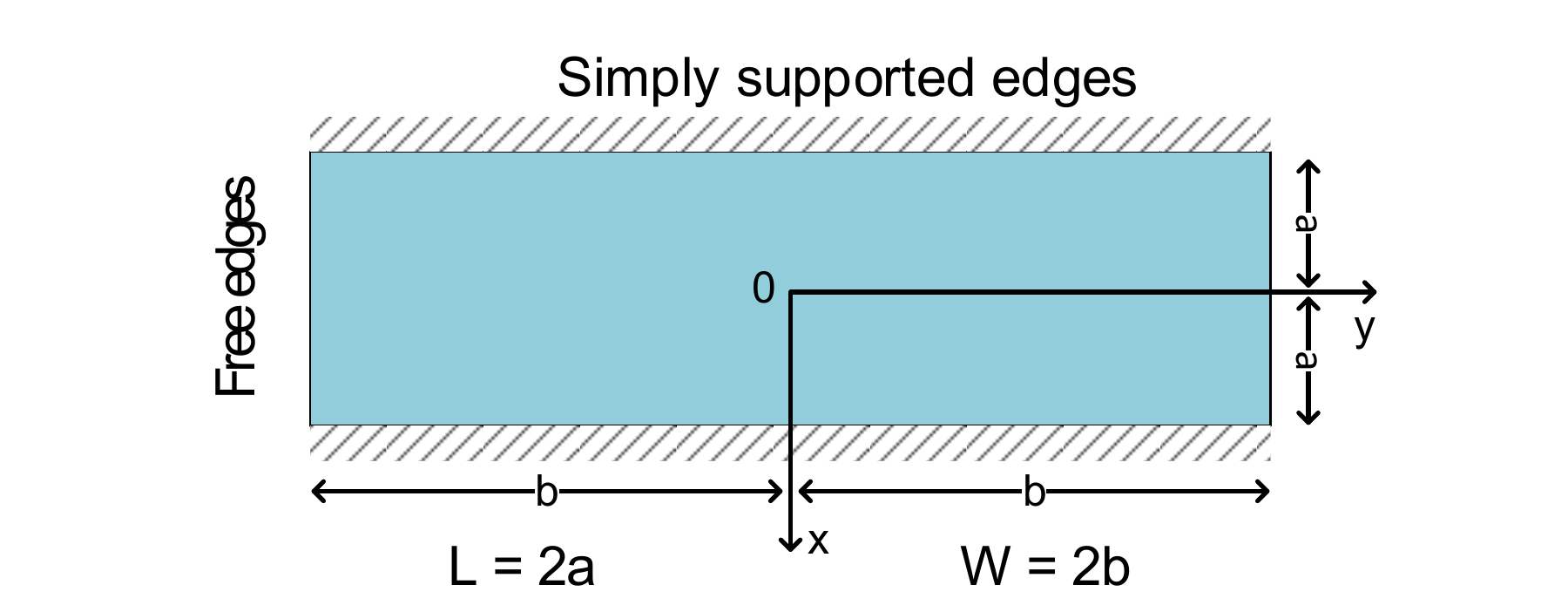}
	\caption{Geometry of the suspended membranes.} 
	\label{fig_S8}
\end{figure}

The measured force deflection relation was linear up to $\sim 10 nm$, which suggested the membranes were dominated by pretension. For deflections higher than 10 nm, the relation was non-linear as the membranes transitioned to the stretching behavior regime.  There is no exact closed form model describing the non-linear (stretching) suspended rectangular membranes loaded at the centre due to the complexity of the geometry and the load, exact solutions can be only found for circular membranes under certain condition due to the axisymmetry of their geometry\cite{Komaragiri_2005,Zhang_2014}. Therefore to fit the acquired data an approximate model was developed. The model was developed for rectangles with width larger than twice their length. This geometry is the most representative of our suspended strips, also the width of the rectangles allows us to neglect the deflection of the free edges. Thus all edges were assumed to be simply supported and immovable. A half cosine deflection profile was assumed. Figure \ref{fig_S8} illustrates the geometry of the membranes under consideration.
The virtual displacement method described in reference \cite{Timoshenko} was used to find an approximate force deflection relation in the non-linear membrane domain (large deflection). After each derivation step the result was compared to the values found in the reference for the square limit. The approximate expressions for the displacements are:
\begin{equation}\nonumber
\omega = d cos \frac{\pi x}{2a} cos \frac{\pi y}{2b}, u = c_1 sin \frac{\pi x}{a} cos \frac{\pi y}{2b}, v = c_2 sin \frac{\pi y}{b} cos \frac{\pi x}{2a} ,
\label{eq:displacements}
\end{equation}
where $\omega$, $u$, $v$ are the displacements in $z$, $x$, and $y$ directions respectively. $d$ is the deflection at the centre of membrane, and $c_1$ and $c_2$ are the maximum displacements in $x$ and $y $directions respectively . The final force deflection relation is given by:

\begin{equation}
F = 4 Et d^3 \dfrac{0.44 a^{12} + 16.3 a^{10}b^{2}+151 a^8 b^4 +3.6 a^6 b^6 + 151 a^4 b^8 + 16.3 a^2 b^{10} +0.44 b^{12}}{a^3 b^3 (a^4+20.5 a^2 b^2 + b^4)^2}.
\label{eq:full}
\end{equation}
where F is the applied force, E is Young's modulus,  and t is the thickness of the membrane. This relation was calculated for a Poisson's ratio of 0.141. For a square, the force deflection relation is:

\begin{equation}
F = 2.7 \dfrac{Et d^3}{a^2}.
\label{eq:square}
\end{equation}
when this relation is adjusted for a uniform load and a Poisson's ratio of 0.25 the relation becomes $q=1.9 Etd^3/a^4$, which agrees with reference \cite{Timoshenko}.


\end{document}